\documentstyle[aps,prb,multicol]{revtex}
\begin{document}
\preprint{to be submitted to Phys. Rev. }
\draft
\title{
Subdecoherent Information Encoding \\
in a Quantum-Dot Array
 }
\author{Paolo Zanardi$^{1,2}$ and Fausto Rossi$^{1,3}$}
\address{
$^{1}$ Istituto Nazionale per la Fisica della Materia (INFM) \\[1ex]
$^{2}$ Institute for Scientific Interchange  Foundation, \\Villa Gualino,
Viale Settimio Severo 65, I-10133 Torino, Italy\\[1ex]
$^{3}$ Dipartimento di Fisica, 
Universit\`a di Modena, 
Via G. Campi 213/A, 
I-41100 Modena, Italy
}
\date{\today}
\maketitle
\begin{abstract}

A potential implementation of quantum-information schemes in 
semiconductor nanostructures is studied. 
To this end, the formal theory of quantum encoding for avoiding errors
is recalled and 
the existence of noiseless states for model systems is discussed.
Based on this theoretical framework, we analyze the possibility of 
designing noiseless quantum codes in realistic semiconductor
structures.
In the specific implementation considered,
information is encoded in the lowest energy sector
of charge excitations of a linear array of quantum dots. 
The decoherence channel considered is 
electron-phonon coupling 
We show that besides the well-known phonon bottleneck, reducing
single-qubit decoherence, suitable many-qubit initial preparation 
as well as register design
may enhance the decoherence time by several orders of magnitude.  
This behaviour stems from the effective one-dimensional 
character of the phononic environment 
in the relevant region of physical parameters.

\end{abstract}
\begin{multicols}{2}[]
\narrowtext
\section{Introduction}\label{s:Int}

Devices using unique quantum-mechanical features 
can perform information processing in a much more efficient
---or even unattainable--- way than those relying just on classical physics.
This fundamental discovery has stimulated in the last few years a big deal 
of work and scientific debates in the new born field of Quantum 
Computation.\cite{QC}
From a conceptual point of view these results represent a serious
challenge to the time-honored notion of  {\em universal computational 
schemes}
independent of an underlying physical theory:
 information as well as computation are {\em intrinsically} physical.
On the other hand, physical realizations of a Quantum Computer
would result  in tremendous practical advantages.

The key ingredients which endow QC devices
with  computational capabilities that supersede their classical counterparts
are basically:
(i) the linear structure of their state space;
(ii) the unitary character of their dynamical evolution;
(iii) the tensorised form of multiparticle state spaces.
The first two properties allow for a parallel processing
of an arbitrary number of data sets, encoded in suitable quantum states. 
By resorting to quantum interference, between different computational
branches, one can selectively
amplify desired parts of the state vector in order to optimize
the probability that  a final (i.e., read-out) measurement will give us
the information we were looking for.
Point (iii) represents another striking departure
from classicality: due to {\em entanglement}, combining different 
quantum systems results in an exponential growth of the available
coding space;  moreover, the tensor-product structure
is at the very basis of many efficient quantum manipulations.

Unfortunately, all this holds just for {\em closed} quantum systems.
Real systems are unavoidably coupled with environmental 
(i.e., non computational)
degrees of freedom. Such open character spoils points (i) and (ii)
eventually turning  quantum computing to classical.
Different computational branches get entangled with different
(quasi-orthogonal) quantum states of the environment 
and their interference is then no longer observable.
From a mathematical point of view,
the relevant state space, given by density matrices,
 has now a {\em convex} structure and 
the allowed quantum dynamics is
described by CP-maps.\cite{KRA} 
Initial pure preparations are typically corrupted on 
extremely short time-scales due to quantum-coherence loss 
that makes them mixed:
the initial information irreversibly leaks out from the system into the huge
number of uncontrollable degrees of freedom of the environment.
This phenomenon ---the so called decoherence problem in QC~\cite{DECO}---
represents the major obstacle for the experimental realization of 
any quantum-computing system.
Other challenging requirements are of course given by the necessity
of being able to perform on a system,   with a {\em well-defined}
state space,
long coherent quantum manipulations ({\sl gating}),
 precise quantum-state synthesis and detection as well.

A major  theoretical achievement has been made by showing
that one can, in principle,   {\em actively} stabilize quantum states
by  means of Quantum Error Correction.\cite{ERROR}
The latter, built in analogy with its classical counterpart,
assumes that the quantum bits ({\em qubits}) are coupled
to independent environments.
The information is then encoded in a subtle redundant way that
allows, monitoring the systems and conditionally carrying on suitable
quantum operations, to tolerate a certain (small) amount of decoherence
and imperfect gating as well.\cite{FAULT}

It is basically the need of dealing with systems sufficiently
decoupled from the external environment that, 
up-to-now, 
has limited the existing realizations to atomic and molecular implementations.
Furthermore, the extremely advanced  technological state-of-the-art in these
fields  allows for the manipulations required in simple QC's.\cite{IMPLE}
However, any interesting QC would require a large number
of quantum gates and qubits as well, and all the present approaches
suffer from the problem of {\em scalability} to  large, i.e. highly integrated,
quantum processors.

One is then naturally led to consider the viability of  solid-state
implementations. 
In particular, by resorting to present semiconductor
technology, one might  benefit synergetically from the
recent progress in ultrafast optoelectronics~\cite{CCC,Shah,Kuhn}
and in nanostructure fabrication and characterization.\cite{QD-reviews}

The first drawback of such a kind of proposal is that the {\em typical}
decoherence time $\tau_D$ in semiconductors is of the order of picoseconds.
On the other hand, the relevant parameter is the ratio between
the typical time-scale of gating $\tau_G$ and $\tau_D$.
Roughly speaking, $\tau_D/\tau_G$ represents the number of 
elementary (coherent) operations
that one could perform on the system before  its coherence being lost.

DiVincenzo and Loss~\cite{DILO} have proposed to use non-equilibrium
spin dynamics in quantum dots for quantum computation.
This exploits  the  low decoherence of spin degrees of
freedom in comparison to the one of charge excitations, 
being the former much less 
coupled with the environment.
Nevertheless, the required magnetic gating is extremely challenging from 
a technological point of view, and the ratio $\tau_D/\tau_G$ does not allow
for the number of gate operations within the decoherence time
required by concrete QC's.

Ultrafast laser technology  is now able to generate electronic excitations
 on a sub-picosecond time-scale
and to perform
on such states a variety of coherent-carrier-control operations.\cite{CCC}
If one can speculate to resort to such a technology for realizing  gating
of {\em charge} degrees of freedom
then  coherence times on nano/microsecond scales
can be regarded as ``long'' ones.

In this paper we analyse in a detailed way the recent idea 
of implementing {\em Quantum Error Avoiding} strategies.\cite{ZR}
The goal here is 
 to suppress decoherence in a quantum register realized
by the lowest energy  {\em charge} excitations of a semiconductor 
quantum-dot array.\cite{ZARO}
In this case, the noise source is given by 
electron-phonon scattering, which is recognized to be the most
efficient    decoherence channel in such a system \cite{Shah,Kuhn}.

Despite of the {\em a priori} complexity
of the three-dimensional (3D) phononic environment, 
we will show that the underlying dynamical-symmetry allows, 
by  means of a proper quantum encoding, to  increase
the decoherence time by several orders of magnitude with respect to the 
bulk value.
The focus of the present paper is mostly conceptual
and the problem of actual preparation/manipulation of the resulting
codewords  will not be addressed.

The paper is organized as follows.
In Sect.~\ref{s:Theory} the formal theory of subdecoherent quantum encoding
is presented and discussed.
Section \ref{s:Appl} deals with the application of the proposed 
subdecoherence theory to realistic semiconductor-based nanostructures; 
More specifically, we will choose as quantum register an array of 
semiconductor quantum dots and for this particular system we will study the
potential sources of decoherence.
In Sect.~\ref{s:Sim} we shall present a detailed investigation of 
decoherence in our quantum-dot array. 
In addition to a short-time analysis, we will present time-dependent 
simulations corresponding to a numerical solution of the Master equation. 
They will show that by means of a proper  initial 
many-electron state preparation
it is possible to extend the carrier-phonon decoherence 
time up to the $\mu$s scale.
Finally, in Sect.~\ref{s:Conc} we will summarise and draw some conclusions.
Appendix A is devoted to a formal analysis of the so-called Circular Model,
which will turn out to play a major role in the semiconductor-based
implementation considered.

\section{Theory of Subdecoherent Quantum Encoding}\label{s:Theory}

In this section we recall the basics of the theory of {\em Noiseless 
Coding}~\cite{ZR} 
in the framework of a Master Equation  (ME) formalism, 
for the register subdynamics.\cite{ZAme}
Generally speaking,
these strategies for preserving quantum coherence
rely on the possibility to  design  an open quantum system $\cal R$ in
such a way that
i) the environment $\cal E$
is effectively coupled only with a subset of the degrees of freedom of $\cal R.$
Information is then encoded in the portion $\cal C$ of Hilbert space spanned
by the remaining (decoupled) degrees of freedom, ii) The environment is coupled
to subset of states $\cal C$ in a state independent fashion.
In both cases  $\cal E$ is not able to extract information from $\cal C:$
the quantum coherence is then {\em passively} stabilized.
From the above points it should be clear the first and major departure
from the Error Correction paradigm: here one assumes the
environmental noise to be correlated.
{\em $\cal E$ is coupled, in a strongly state-dependent way, with  collective
states of $S.$} 

 Before embarking in  a detailed analysis of sub-decoherence 
let us  shortly discuss two very simple examples, that show
how this notion can come about.

i)  Let us consider $N$ isospectral linear oscillators $H_{\cal R}=\omega \sum_{j=1}^N
b_j^\dagger\,b_j$
coupled with the vacuum fluctuations i.e., zero temperature, of a bosonic field $a_k$
by an Hamiltonian of the form 
$H_{\cal I}= \sum_{jk} (g_{kj}  a_k^\dagger\,b_j +\mbox{h.c.}).$
Suppose now that $g_{kj}=g_k\, \forall j.$
By introducing the Fourier transformed operators $b_q\equiv 1/\sqrt{N}\sum_j e^{i\,q\,j} b_j$
(bosons as well) one immediately sees that only the zero-modes are 
actually coupled:
$H_{\cal I}= b_0^\dagger (\sum_k g_k a_k)+\mbox{h.c.}$ and 
$H_{\cal R}=\omega \sum_q  b_q^\dagger\,b_q.$
Therefore, any state of the (infinite-dimensional) subspace
\begin{equation}
{\cal C}= |0\rangle_0\bigotimes_{q>0} {\cal H}_q
\end{equation}
will evolve unaffected by the environment in that 
$ H_{\cal I}\,{\cal C}\otimes |0\rangle_{\cal E}=0.$

ii) Let the system-environment interaction Hamiltonian be of the form
$H_{\cal I}=\sum_\mu R_\mu\otimes E_\mu, $ where $X_\mu\in\mbox{End}\,{\cal H}_{\cal X}\,
({\cal X}={\cal R},\,{\cal E})$. 
Moreover, let us suppose that 
the Hermitian $R_\mu$'s  are commuting operators, 
i.e., they span an {\em abelian}
algebra $\cal A$. Let ${\cal C}\subset {\cal H}_{\cal R}$ a simultaneous 
eigenspace of $\cal A.$ This means that 
\begin{equation}
H_{\cal I}|_{{\cal C}}=\sum_\mu \rho_\mu\otimes E_\mu \quad (\rho_\mu\in{\bf{R}}),
\end{equation}
in other words, if one restricts himself to ${\cal C}$ the interaction with 
the environment amounts simply to a {\em state-independent 
renormalization of } $H_{\cal E}.$
It is then clear that ---provided ${\cal C}$ is invariant under the system 
self-Hamiltonian
$H_{\cal R}$---   
any initial preparation in ${\cal C}$ evolves in a unitary fashion
regardless the strength of the system-environment coupling and 
the environment initial state as well. 
Of course, for all this to be useful in quantum encoding one must have
$\mbox{dim}\,{\cal C}>1.$ 

\subsection{Master-equation approach}\label{s:ME}

The system under investigation $\cal R$ is given by $N$ {\em identical} 
two-level systems 
({\em $N$-qubits quantum register }),
 representing
our computational degrees of freedom, coupled with an external 
(uncontrollable) environment.
The register $\cal R$  will be described in the spin $1/2$ language by means of the usual Pauli spin matrices
 $\{ \sigma_i^z\,\,\sigma^{\pm}_i\}_{i=1}^N $ generating $N$ {\em local} $sl(2)$ algebras
\begin{equation}
[\sigma^+_i,\,\sigma^-_j]= 2\,\delta_{ij}\sigma_i^z,\quad
[\sigma^z_i,\,\sigma^{\pm}_j]= \pm\delta_{ij} \sigma^{\pm}_i.
\label{sl2}
\end{equation}
The collective spin operators $S^\alpha=\sum_{i=1}^N\sigma^\alpha_i,\,(\alpha=\pm,z)$
span a $sl(2)$ algebra as well, it will be referred to as the {
\em global} $sl(2).$
The environment $\cal E$ will be described by a set of non interacting 
harmonic oscillators
with bosonic field operators 
$[b_k^\dagger,\,b_{k'}]=\delta_{kk'}.$

The total Hamiltonian is assumed to be $H=H_{\cal R}+H_{\cal E}+ H_{\cal I},$
where $H_{\cal R}=E \,S^z$ and $H_{\cal E}=\sum_k \omega_k\, b_k^\dagger\,b_k$
are, respectively, the register and the environment self-Hamiltonians. 
Here, $E$ represents the energy spacing between  
states $|0\rangle_i$ and $|1\rangle_i$
in each qubit.
The ${\cal R}-{\cal E}$  interaction is given by 
\begin{equation}
H_{\cal I}= \sum_{ki}( g_{ki} \,b^\dagger_k\,\sigma^-_i +\mbox{h. c}).\label{Hint}
\end{equation}
Let us now briefly recall  the standard Born-Markov scheme
for tracing out the ${\cal E}$ degrees of freedom and 
obtaining  a Master equation for the register subdynamics.
The Liouville-von Neumann equation for the 
{\em total} density matrix of ${\cal R}\otimes{\cal E}$
in the interaction picture
reads
$i\,\partial_t\tilde\rho =[H_{\cal I},\,\tilde\rho]$.
 One assumes a factorized initial condition
$\tilde\rho(0)=\rho\otimes\Omega.$
After a formal time integration one obtains 
\begin{eqnarray}
\tilde\rho(t)&=& \tilde \rho(0)+
\int_0^t d\tau [H_{\cal I}(\tau),\tilde\rho(\tau)]
\nonumber \\
&=& \tilde \rho(0) - i\int_0^t d\tau [H_{\cal I}(\tau),\tilde\rho(0)]
\nonumber \\
&+&(-i)^2
\int_0^t d\tau \int_0^\tau d\tau^\prime 
[H_{\cal I}(\tau),[H_{\cal I}(\tau^\prime),\tilde\rho(\tau^\prime)]]
\end{eqnarray}
Now we set $\tilde\rho(\tau^\prime)=\rho(\tau^\prime)\otimes \Omega$ 
($ \Omega \sim e^{-\beta H_{\cal E}}$ )
and we perform a partial trace over ${\cal E}$ 
in order to get  an equation
for the reduced density matrix of $\cal R$: 
$\rho(t)=\mbox{tr}^{\cal E} \tilde \rho(t).$
The resulting ME is of the form 
$\dot \rho ={\cal L}(\rho).$ The  Liouvillian {\em superoperator}
${\cal L}$ is given by the sum of two contributions:
${\cal L}_u$ representing the unitary component of the dynamics ruled
by the (renormalized) register self-Hamiltonian; 
${\cal L}_d$
describing the irreversible decoherence/dissipation processes induced
by the coupling with the external bath.
By denoting with $H_{\cal R}$ the unperturbed register self-Hamiltonian,
one has that 
${\cal L}_u (\rho)={i/\hbar}\,[\rho, { H}_{\cal R}+\delta{ H}_{\cal R}]$
where the environment-induced
$\delta{ H}_{\cal R}$  is given by
\begin{equation}
\delta{ H}_{\cal R}
 =\sum_{\eta=\pm}
\sum_{ ij=1}^N \Delta_{ij}^{(\eta)} \sigma^{-\eta}_i\,\sigma^{\eta}_{j}
\label{Lunit}.
\end{equation}
These contributions ---usually referred to as the Lamb-shift terms---
describe a sort of qubit-qubit effective interaction mediated by the 
external environment. 
The dissipative Liouvillian is given by 
${\cal L}_d=\sum_{\eta=\pm}{\cal L}^{\eta}_d,$
where
\begin{equation}
{\cal L}^{\eta}_d(\rho)=\frac{1}{2\,\hbar}\sum_{ij=1}^N
\Gamma^{(\eta)}_{ij}\, \left (
[\sigma_i^\eta
\,\rho,\,\sigma_{j}^{-\eta}]+
[\sigma_i^\eta,\,\rho\,\sigma_{j}^{-\eta}]\right ),
\label{Ld}
\end{equation}
Here, the term $\eta=-$ ($\eta=+$) is associated 
to deexcitation (excitation)
processes of the qubits by emission (absorption) of bosonic quanta.
The Hermitian matrices  ${\bf \Gamma}$ and ${\bf \Delta}$ 
are the input  data defining  our ME, their actual form
depends on the details of the physical constants 
($E,\,\{\omega_k\}_k\, \{g_{ki}\}$, etc.)
and will be given later.

As far as the analysis of this section is concerned is sufficient to know that
${\bf \Gamma}\ge 0.$
One can go on with general considerations
by diagonalizing   ${\bf{\Gamma}}^{(\eta)}$
in order to  obtain the  canonical form for the dissipative part of the 
Liouvillian~\cite{LIN}
\begin{equation}
{\cal L}_d(\rho)= \frac{1}{2\,\hbar}\sum_{\eta=\pm, \mu=1 }^N \lambda_\mu^\eta\left (
[L_\mu^\eta\,\rho,\, L_\mu^{-\eta}]+ [L_\mu^\eta,\,\rho\, L_\mu^{-\eta}]\right ),
\label{lindblad}
\end{equation}
where $\{\lambda_\mu^\eta\}$ are the (non-negative) eigenvalues of $
{\bf{\Gamma}}^{(\eta)}.$
Moreover, $L_\mu^\eta=\sum_i u^\mu_i\,\sigma_i^\eta,$ 
$u_i^\mu$ denoting the components of the eigenvectors
of ${\bf{\Gamma}}^{(\sigma)}.$
The $L_\mu^\eta$'s will be referred to as the Lindblad operators.
The  operator  (Lie) algebra $\cal A$ spanned by the  Lindblad operators
contains the information about the existence of 
coding spaces  stable at least on a short-time-scale.
The finite-time stability depends on the interplay between the dissipative 
and the unitary components
of the Liouvillian 
in ${\cal H}_{\cal R}.$

In order to quantify the efficiency of the environment in destroying
quantum coherence it is  useful 
to  define a {\em (first-order) decoherence time (rate) $\tau_1$ 
($\tau_1^{-1}$) }
by means of the short-time expansion of the {\em fidelity}~\cite{ZAme}
for {\em pure} initial state preparations $|\psi\rangle$
\begin{equation}
F(t)\equiv \langle \psi|
\, \rho(t)\,|\psi\rangle = 1- \frac{t}{\tau_1} +o(t^2).
\end{equation}
From Eq.~(\ref{lindblad}) one obtains
\begin{equation}
\tau_1^{-1}[|\psi\rangle]= \sum_{\eta=\pm, \mu=1}^N \lambda_\mu^\eta \left (
\| L_\mu^\eta\,|\psi\rangle\|^2 -|\langle\psi|\,L_\mu^\eta\,|\psi\rangle|^2
\right ).
\label{tau_1}
\end{equation}
This expression is nothing but a sort of fluctuation-dissipation relation
connecting the dispersion of the Lindblad operators $L_\mu^\eta$
in the initial register state with the rate at which 
quantum coherence is destroyed.
It is important to point out that the unitary   component of the Liouvillian 
does not contribute
to the first-order decoherence time.
If $\tau_1^{-1}[|\psi\rangle]=0$ then the state $|\psi\rangle$ will be 
called {\em subdecoherent}
and  a linear subspace ${\cal C}\subset {\cal H}_{\cal R}$ will be 
referred to as a {\em subdecoherent code.}

In general, the register Hilbert space splits  in 
${\cal A}$-invariant subspaces,
\begin{equation}
{\cal H}_{\cal R} =\oplus_{J }\oplus_{r=1}^{n_J} {\cal H}^{(J)}_r,
\label{split}
\end{equation}
where  $J$ labels the irreducible representations (irrep)  of $\cal A,$ and the integers
$n_J$ are the associated multiplicities [$ {\cal H}^{(J)}_r\cong  {\cal H}^{(J)}_{r^\prime}$.]   
 The {\em singlet sector} $\cal C$ of $\cal A$ is the direct sum 
(possibly empty) of the one-dimensional
irreps.
In Ref.~\cite{ZAme} it has been shown that, for  non-abelian $\cal A,$
the sub-decoherent codes coincide with $\cal C.$
In an equivalent group-theoretic language one can say that
the code $\cal C$ is the { subspace} of vectors {\em invariant} under the 
action of group
${\cal G}=\exp\, {\cal A}$ generated (infinitesimally) by the Lindblad operators $L_\mu.$
[$\cal C$ is the trivial $\cal G$-representation space.]
This group acts, of course, on the general mixed states: $\rho\mapsto X\,\rho\,X^\dagger,\;
(X\in{\cal G}).$ 
The same argument holds for the subdecoherent (pure) states.
When $\cal C$ is invariant under the action of $H_{\cal R}^\prime$;  
then the contribution to the dynamics of ${\cal L}_d$ vanishes
\begin{equation}
\rho=|\psi\rangle\langle\psi|\mapsto e^{-i\,t\,H_{\cal R}^\prime}\,\rho\, e^{i\,t\,H_{\cal R}^\prime}
\quad (\forall |\psi\rangle\in{\cal C},\,t\ge 0)
\end{equation}
The finite-time evolution is unitary, in this case ${\cal C}$   
will be referred to as
 {noiseless code}: {\em quantum coherence is preserved -- in principle --
for an arbitrarily long time.}
When $\cal C$ is not $H_{\cal R}^\prime$-invariant the 
initial preparation $|\psi\rangle$
on a greater time-scale leaks
out from the code  and its quantum coherence will be eventually washed out.
For instance, the  condition   $[H_{\cal R}^\prime,\,{\cal A} ]=0,$  suffices
to have such a noiseless coding or even that $H_{\cal R}^\prime$ 
belonging to the
{\em associative} operator algebra ${\cal A}_{a}$ generated by the $L_\mu$'s
and the identity operator.
Notice that if $\cal C$ is subdecoherent for the $L_\mu$'s 
it is subdecoherent for any set of Lindblad operators included in ${\cal A}_{a}.$

From a physical point of view, the algebra ${\cal A}$
of Lindblad operators represents the set  of the  register modes
that are incoherently excited by the environment; looking for  states
that are annihilated by as many Linbdlad operators as possible
is  then as looking for  states that  are `` vacua ''
for the largest number of such  excitations and therefore
maximally decoupled with environment.\cite{notDeco}
It is important to emphasize that such a decoupling can be achieved
thanks the algebraic-dynamical structure of the model
without any assumptions about the (weakness) of the register-environment 
interaction.
Loosely speaking, one can say that for generic ${\bf \Gamma}$'s, 
the Liouvillian is such that,
given {\em any}
register preparation,  the environment forces the coding system
to  explore the totality of its Hilbert space so that there is no safe place
 where  storing quantum information, instead for some ``magic'' ${\bf \Gamma}$
the Lindblad algebra gets smaller allowing just for a limited probing
of the register space of states by the environment
strongly dependent on the initial register data: 
free room is left for {`` hiding ''}
quantum information.

Rather interestingly, the problem of analysing state 
stability against decoherence 
can be cast in a Hamiltonian form by observing
that,
for an initial condition $|\psi\rangle$ that is a $S^z$-eigenstate one has
$\tau_1^{-1}= 
\langle\psi| \,\tilde H\,|\psi\rangle$ where
\begin{eqnarray}
\tilde H = \sum_{\eta=\pm, \mu=1}^N \lambda_\mu^\eta L_\mu^{-\eta}\, L_\mu^{\eta}
= \sum_{ij=1}^N(\Gamma_{ij}^{(-)} \sigma_i^+\sigma_j^- +\Gamma_{ij}^{(+)}\,\sigma_i^-\sigma_j^+)
\label{Ham}
\end{eqnarray}
 In other words: {\em the problem of finding  decoherence rates is mapped onto
the  spectral problem
 for the (positive) operator (\ref{Ham}).}
In particular, ``robust'' states (i.e., the ones with minimal decoherence rates)
are {\em ground states} of $\tilde H.$ 
Let $E_N$
denote the lowest eigenvalue of $\tilde H.$
$E_N=0$ means that there exist sub-decoherent states,
in this case  ${\cal C}\equiv \mbox{Ker} \,\tilde H$ and
$d_N\equiv \mbox{dim Ker} \,\tilde H$  gives the dimension
of the code.
The sub-decoherence property is stable against small perturbations of the state. 
 Indeed if $|\psi\rangle\in\mbox{Ker}\,\tilde H\mapsto |\psi\rangle+ |\delta\psi\rangle$
then $\delta \tau_1^{-1} =\langle\delta\psi|\,H\,|\delta\psi\rangle\ge 0$

\subsection{A Simple Example}

To better illustrate the situation let us consider the  $N=2$ case.
The model  (\ref{Ham})  is  soluble in elementary way \cite{notN2}.
We assume $\Gamma_{11}^{(\pm)}=\Gamma_{22}^{(\pm)}\equiv\Gamma^{(\pm)}$
and $\Gamma_{12}^{(\pm)}=\Gamma_{21}^{(\pm)}=\Gamma^{(\pm)}\,\beta,$ moreover $\Gamma^{(-)}\ge\Gamma^{(+)}.$
From positivity it follows that $|\beta|\le 1.$
The spectrum is given by
\begin{eqnarray}
E_{11} &=& 2\,\Gamma^{(-)},\quad E_{00}= 2\,\Gamma^{(+)},
\nonumber \\
E_{t,\,s}&=& (\Gamma^{(-)}+\Gamma^{(+)})(1\pm \beta),
\end{eqnarray}
with eigenstates given respectively by $$|11\rangle,\,|00\rangle,\,
2^{-1/2}(|01\rangle\pm|10\rangle).$$
If $\Gamma^{(+)}>0,$ for $|\beta|\le ( \Gamma^{(-)}- \Gamma^{(+)})( \Gamma^{(-)}+ \Gamma^{(+)})^{-1}\equiv \beta_c$
one has $E_2=E_{00},$ for $\beta >\beta_c$ ($\beta <\beta_c$ one finds $E_0=E_s$ ($E_0=E_t.$)
 $\Gamma^{(+)}=0\Rightarrow E_2=E_{00}=0.$ Finally for $\beta=\pm 1$ 
one has again $E_2=0,$
with eigenstates given  by $|\psi_{s,t}\rangle.$
In summary, subdecoherent states exist in a  subset of the boundary of the 
$\bf {\Gamma}$ manifold.
This result is quite general: for {\em generic} $\bf {\Gamma}$'s 
one has $E_N[{\bf{\Gamma}}]>0,$
the subdecoherence condition $E_N[{\bf{\Gamma}}]=0$ is 
fulfilled just in a ``zero-measure'' 
set of the Hamiltonian models (\ref{Ham}).
Of course this is simply due to the fact that for  a generic $\bf {\Gamma}$'s
gives rise to a Lindblad algebra $\cal A$ is too large for
admitting a (non-trivial) singlet sector.
 
Turning back to the general $N$ case,
to exemplify the collective nature of the decoherence-dissipation dynamics
let us consider 
the states ($N$ {\em even})
\begin{eqnarray}
|\psi_{sym}\rangle &\equiv& (S^\dagger)^{N/2}\,|{\bf{0}}\rangle,\nonumber \\
|{\cal D}\,\gamma\rangle &\equiv& \otimes_{(i,j)\in{\cal D}}( |01\rangle-(-1)^{\gamma(i,\,j)}|10\rangle)_{ij},
\label{States}
\end{eqnarray}
where ${\cal D}$ is a dimer partition of the qubit array, and $\gamma\colon {\cal D}\rightarrow \{0,\,1\}.$
The first state in (\ref{States}) is simply the totally symmetric
$S^z=0$ state (belonging to the $sl(2)$ multiplet of the vacuum)
whereas the  $|{\cal D}\,\gamma\rangle$'s are products 
of singlet or triplet pair-states depending on the signature $\gamma$ 
of the register dimer partition ${\cal D}.$ 
This latter family of states (\ref{States}) will play an important role 
in the following.
Notice that, for $\gamma={\bf 0},$ one gets
global $sl(2)$ singlets corresponding to zero total angular momentum $S^2.$
In terms of Hadamard transformations and controlled-not operators the $|{\cal D}\,\sigma\rangle$'s
can be {\em sinthetized}  as follows from a pure product state
\begin{equation}
|{\cal D} \gamma\rangle= \otimes_{(l,m)\in{\cal D}}{\tt{cnot}}_{lm}
 H_l\,| \gamma(l,m)+1,1\rangle_{lm}
\end{equation}
With a straightforward calculations one finds that the first-order decoherence rates of states 
(\ref{States})
are given respectively by $(\tau_\alpha/\tau_0)^{-1}= f_\alpha({\bf{\Gamma}}),\,(\alpha=sym,{\cal D}\gamma)$
in which $\tau_0^{-1}=\Gamma_0\,N/2 $
 is the decoherence rate for uncorrelated qubits and
($\tilde \Gamma_{ij}\equiv \Gamma_{ij}/\Gamma_0$)
\begin{eqnarray}
f_{sym}&=&  1+\frac{1}{N-1}\Re\sum_{i<j}
 \tilde\Gamma_{ij} \nonumber \\
f_{{\cal D}\,\gamma}&=& 1-\frac{2}{ N} \Re\sum_{(i,j)\in{\cal D}}(-1)^{\sigma(i,\,j)} \tilde \Gamma_{ij}
\label{rate}
\end{eqnarray}
where $\Gamma_0=\Gamma_{ii},\,(i=1,\ldots,N)$
The $f_\alpha$'s contain the information about the degree of many-qubit 
correlation
in the decay process:
if ${\bf{\Gamma}}\propto {\bf{I}}$ one has $f_\alpha=1$ the qubits decohere
independently.

\section{Application to Semiconductor Nanostructures}\label{s:Appl}

In this section we shall discuss a potential application of the above 
sub-decoherent quantum-encoding strategies to realistic, 
i.e., state-of-the-art, semiconductor-based nanostructures.
Since in semiconductors the primary source of decoherence is known to be 
carrier-phonon scattering, we will consider as prototypical systems 
quasi zero-dimensional (0D) structures, for which the reduced phase-space 
available allows for a significant suppression of phonon-induced energy 
relaxation and dephasing.

We will choose as prototype of quantum register an array of 
semiconductor quantum dots. 
In particular, we will consider as quantum dot (QD) a GaAs/AlGaAs structure 
similar to that studied in \onlinecite{QD-str}.
Here, various effects due to carrier-carrier interaction 
will {\em not} be considered. 
This is, of course, a potential limitation of our analysis,
especially in relation to state preparation/manipulation (not addressed in 
this paper).
Indeed, the latter requires a controllable source of entanglement, 
i.e., a qubit-qubit interaction that might be provided by  
``switchable'' Coulomb couplings.\cite{BEDJ}
On the other hand, our coding states will involve
{\em single-electron} occupations only;
For such states the intra-dot Coulomb repulsion is clearly absent,
while the inter-dot one at the  distances 
relevant for our quantum encoding
is found to be negligible.\cite{MMR1}
Moreover, 
since the system under consideration is based on intrinsic III-V materials,
carrier-impurity scattering is negligible.

Generally speaking, 
 Hamiltonian modifications will result in leakage
from the coding subspace  only on a longer time-scale
with respect to the phonon-scattering one, 
i.e., it does not affect
the stability classification based on $\tau_1$ (see Sect.~\ref{s:Theory}).
Finally, we would like to stress that there exists
a whole class of interactions leaving the code invariant.\cite{ZR}

\subsection{Free-Carrier States in the Quantum-Dot Array}

The confinement potential $V^{\rm 0D}$ giving rise to the 
quasi-0D carrier states in such a QD structure
is properly described in terms of a quantum-well (QW) profile 
$V^\parallel$ 
along the growth direction of the structure plus a two-dimensional (2D) 
parabolic potential 
$V^\perp$ 
in the normal plane.
More specifically, a carrier within the $i$-th QD structure is described by the 
following single-particle Hamiltonian 
\begin{eqnarray}
h_i & =& 
-{\hbar^2\nabla_{\bf r}^2\over 2m^*} + V^{\rm 0D}({\bf r}) = 
\left(
-{\hbar^2\nabla_{{\bf r}_\perp}^2\over 2m^*} + V^\perp({\bf r}_\perp)
\right) \nonumber \\ &+&
\left(
-{\hbar^2\nabla_{{\bf r}_\parallel}^2\over 2m^*} + 
V_i^\parallel({\bf r}_\parallel)
\right) = 
H^\perp + H^\parallel\ ,\label{singleH}
\end{eqnarray}
where
\begin{eqnarray}
  V^\perp({\bf r}_\perp)  = \frac{1}{2} m^* \omega^2 |{\bf r}_\perp|^2,\, 
\end{eqnarray}
is the 2D harmonic-oscillator potential in the ($x,y$) plane 
perpendicular to the ($z$) array axis (which coincides with the growth 
axis of the QD structure), while 
$V^\parallel_i({\bf r}_\parallel)$ is 
a 1D square-well potential centered at 
${\bf r}_\parallel^i = i\,a\,{\bf{\hat z}}$ with 
width $d$ and infinite walls,\cite{notWA} 
$a$ being the array periodicity, i.e., the 
inter-dot distance.
This choice for the single-particle Hamiltonian, 
even though not generally valid, well
describes the 0D carrier confinement of the low-energy states in the QD 
structure, 
which are the only relevant states for the quantum encoding considered.
We would like to point out that the very same QD model
turned out to be able to explain, in a quantitative way,
the addition spectra reported in \onlinecite{QD-str}.\cite{MMR}

The  Hamiltonian (\ref{singleH} )is elementary soluble, 
its spectrum being the sum 
of the parallel and perpendicular contributions:
\begin{equation}
\epsilon_{n\nu} = E^\perp_n + E^\parallel_\nu =
(n_x+n_y+1)\hbar\omega + {\pi^2\hbar^2 \nu^2\over 2m^* d^2}\ .
\end{equation}
The corresponding 3D eigenstates will be factorized according to:
\begin{equation}
\phi_{i,n\nu}({\bf r}) = \phi^\perp_{n_x,n_y}({\bf r}_\perp) 
\phi^\parallel_\nu({\bf r}_\parallel-i\,a)\ .
\end{equation}
The total free-carrier Hamiltonian describing our QD array can then 
be expressed  in the (second-
quantized) form
\begin{equation}
H_{\cal R}=\sum_{i,\alpha}\epsilon_\alpha c_{i\alpha}^\dagger\,c_{i\alpha},
\label{Harray}
\end{equation}
where the fermionic operators $ c_{i\alpha}^\dagger$ ( $c_{i\alpha}$)
create (destroy) an electron in the $i$-th QD in state 
$\alpha \equiv n_x n_y \nu$.

\subsection{Carrier-Phonon Coupling}

The Hamiltonian describing the free phonons of a semiconductor crystal
is given by~\cite{IF-Ph}
\begin{equation}
{ H}_{\cal E} = \sum_{\lambda{\bf q}} \hbar\omega_{\lambda{\bf q}}\,
b^\dagger_{\lambda{\bf q}} b^{ }_{\lambda{\bf q}}
\end{equation}
where $\lambda$ and ${\bf q}$ denote,
respectively,  the phonon mode (e.g. acoustic, optical, etc) and the phonon 
wavevector.

The coupling of phonons with the electrons in the QD array is described by 
the following carrier-phonon interaction Hamiltonian:
\begin{equation}
{ H}_{\cal I} = \sum_{i\alpha,i'\alpha';\lambda{\bf q}}
\left[
g^{ }_{i\alpha,i'\alpha';\lambda{\bf q}}
c^\dagger_{i\alpha} b^{ }_{\lambda{\bf q}} c^{ }_{i'\alpha'} +
\mbox{h.c.} \right].
\label{H_I}
\end{equation}
Where
\begin{equation}
g_{i\alpha,i'\alpha';\lambda{\bf q}} = \tilde{g}_{\lambda{\bf q}}
\int \phi^*_{i\alpha}({\bf r})
e^{i{\bf q \cdot r}} \phi^{ }_{i'\alpha'}({\bf r}) d{\bf r}
\label{g}
\end{equation}
are the matrix elements of the phonon potential between the
quasi-0D states $i\alpha$ and $i'\alpha'$. The explicit form of the coupling
constant $\tilde{g}_{\lambda{\bf q}}$ depends on the particular phonon mode.

\subsection{The Qubit Register}

In the proposed information-encoding scheme the single {\sl qubit} is given by 
the two lowest energy levels of the QD structure.
Since the width $d$ of the GaAs QW region is typically of the order of few 
nanometers, the energy splitting due to the quantization along the growth 
direction is much 
larger than the confinement energy $\hbar\omega$ induced by the 2D 
parabolic potential $V^\perp$
(typically of a few meV).
Thus, the two lowest-energy states ---state $|0\rangle$ and
$|1\rangle$--- realizing our qubit 
are given by products of 
the QW ground state times the ground or first excited state of the 2D 
parabolic potential \cite{note-degeneracy}. 

More specifically, they are given by
 \begin{eqnarray}
\langle {\bf{r}}|0\rangle_i 
&=&\phi_0^\perp(x)\,\phi_0^\perp(y)\,\phi^\parallel_{i,0}(z),\nonumber \\
\langle {\bf{r}} |1\rangle_i 
&=&\phi_0^\perp(x)\,\phi_1^\perp(y)\,\phi^\parallel_{i,0}(z)
\label{qbit}
\end{eqnarray}
where
\begin{eqnarray}
\phi_0^\perp(x)&=& C_0\,e ^{-a_0\,x^2},\; 
C_0=(2\,a_0/\pi)^{1/4},\; 
a_0=\frac{m^*\,\omega}{2\,\hbar}
\nonumber \\
\phi_1^\perp(x) &=& C_1\,x\,e ^{-a_0\,x^2},\; C_1=2\,a_0^{3/4}(2/\pi)^{1/4}
\end{eqnarray}
 are, respectively, 
the ground and first excited states of the harmonic oscillator in the 
(perpendicular) $xy$ plane,
and 
\begin{equation}
\phi_{i,0}^\parallel (z) = C_z\,\cos [\frac{\pi}{d}(z-i\,a)], \;, 
C_z=\sqrt{2/d}
\label{qbitz}
\end{equation}
is the ground state of the $i$-th quantum-well potential  parallel 
to the array axis ($\phi_{i,0}^\parallel (z)=0$ for $|z-i\,a|\ge d/2$).

Notice that the only dependence on the QD label $i$
of the qubit states is in the  $z$-component of the wavefunction.

Since we are restricting ourselves to the low-energy sector $\alpha=0,\,1$
in the absence of inter-dot ($i \ne i'$) transitions,
the only relevant fermionic bilinears in Eq.~(\ref{H_I})
are given by $X_i=c_{i1}^\dagger c_{i0}$
and their conjugates.
Consistently with the
 commutation relations $[X_i,\,X_j^\dagger]=\delta_{ij}
(n_i^1-n_i^0)\equiv 2\,\sigma_i^z$,
these bilinears
 {\em  can be described  by the spin $1/2$ operators $\sigma_i^{\pm}.$}
Let $|{\bf{0}}\rangle= \prod_{i=1}^N c_{i0}^\dagger\,|\mbox{vac}\rangle$
the reference state built over
the electron vacuum by occupying all the $|0\rangle_i$.
Our {\em reduced} Hilbert space containing the computational degrees of freedom
is then given by
\begin{eqnarray}
{\cal H}_{\cal R}=\mbox{span}\{\prod_{i=1}^N X^{\alpha_i}_i\,|{\bf {0}}\rangle\,|\, \alpha_i=0,\,1\}
\cong \bigotimes_{i=1}^N {\bf{C}}^2
\end{eqnarray}
Any process inducing transitions out of this subspace
will result in a computational error.
Let $\Delta$ being the energy gap between $|1\rangle$
and the higher excited states (in the present case 
$\Delta = \hbar\omega$)
and $T$ the environment (i.e., lattice) temperature;
this sort of {\em leakage} errors occur with low probability as long as
$\Delta \gg k_B\,T.$

By denoting with $E \equiv \epsilon_{i,1}-\epsilon_{i,0} = \hbar\omega$ 
the energy spacing 
between our two qubit levels,
the free-carrier Hamiltonian (\ref{Harray}) for our qubit register, i.e. 
restricted to the low-energy sector $\alpha=0,\,1$,
can then be written as 
\begin{equation}
H_{\cal R} = E \sum_{i=1}^N \sigma_i^z\ ,
\end{equation}
where $\sigma_i^z$ denotes the usual diagonal 
Pauli matrix 
acting on the $i$-th qubit.

Let us now consider again the carrier-phonon interaction Hamiltonian 
(\ref{H_I}).
Within the carrier model considered, wavefunctions corresponding to different 
QD's do not overlap; thus
one has $g_{i\alpha,i'\alpha';\lambda{\bf q}} = 0$ for $i \ne i'$,
i.e., phonons induce intra-dot (intra-qubit) transitions only.
The coupling constants associated to the relevant elementary processes in 
our qubit register are
$g_{i,\lambda {\bf q} }\equiv g_{i1,i0;\lambda {\bf q}}, 
\bar g_{i,\lambda{\bf q}} \equiv g_{i0,i1;\lambda {\bf q}}.$
More specifically, starting from the explicit form of the single-particle 
wavefunctions $\phi$ in (\ref{qbit})
one finds
$g_{i,\lambda{\bf q}} = 
\tilde{g}_{\lambda{\bf q}}\,
g_x({q}_x)\,g_y({q}_y)
\,g_z({q}_z,\,z_i)$
[${\bf{q}} = (q_x,\,q_y,\,q_z)$]
where
\begin{eqnarray}
g_x({q}_x) &=& \langle \phi^\perp_0| e^{i\,q_x\,x} |\phi^\perp_0\rangle=
 \exp (-\frac{q_x^2}{8\,a_0})
\nonumber \\
g_y({q}_y) &=& \langle \phi^\perp_1| e^{i\,q_y\,y}|\phi^\perp_0\rangle=
i\,\frac{1}{2\,a_0^{1/2}} q_y\, \exp(-\frac{q_y^2}{8\,a_0})
\nonumber \\
g_{iz}({q}_z) &=& \langle \phi^\parallel_{i,0}| e^{i\,q_z\,z} |\phi^\parallel_{i,0}\rangle=
\frac{ 8 \pi^2}{d^3\,q_z}\frac{ \sin (q_z\,d/2)}
{ q_0^2-q^2_z}\,e^{i\,q_z\,z_i}
\label{const}
\end{eqnarray}
 and $q_0={2\,\pi}/{d}$

Within these assumptions the carrier-phonon interaction Hamiltonian 
(\ref{H_I}) can be cast in to the form (\ref{Hint}):
\begin{equation}
H_{\cal I}= \sum_{ki}( g_{ki} \,b^\dagger_k\,\sigma^-_i +\mbox{h. c})\ ,
\end{equation}
where the bosonic label $k$ now corresponds to the phonon modes of the 
crystal, i.e.,
$k \equiv \lambda{\bf q}$. 

Following the Born-Markov procedure 
discussed in Sect.~\ref{s:Theory}, 
one finds the following result for the matrices ${\bf{\Gamma}},\,{\bf{\Delta}},$
defining our ME \cite{note-disp}
\begin{eqnarray}
\Gamma^{(\pm)}_{ij}&=&2\,\pi\,\sum_k g_{ki}\,\bar g_{kj}\, (\,n_k +\theta(\mp)\,)
\,\delta(\hbar\omega_k-E),\nonumber \\
\Delta^{(\pm)}_{ij} &=&{\cal P}\sum_k\frac{g_{ki}\,\bar g_{kj}}
{\hbar\omega_k-E}\,(\,n_k +
\theta(\mp)\,)\ .
\label{explicit}
\end{eqnarray}
Here, $\theta$ is the customary Heaviside function, and ${\cal P}$ denotes 
the principal part.
From these relations it follows that ${\bf{\Gamma}}^{(\pm)}$ and ${\bf{\Delta}}^{(\pm)}$
are hermitian as expected.
 Furthermore ${\bf{\Gamma}}^{(\pm)}\ge 0$ 
and 
${\bf{\Gamma}}^{(-)}\ge {\bf{\Gamma}}^{(+)
}$.
Since for the QD structures considered 
the energy splitting $E$ 
is typically much smaller than  the optical-phonon energy [$36\, mev$ in GaAs]
the only phonon modes $k = \lambda{\bf q}$ involved are the acoustic ones.
In this case,
by considering carrier-phonon coupling due to deformation potential,
one has 
$\tilde{g}({\bf q}) = \sqrt{ \frac{ \hbar\varepsilon q^2}{2\,\rho\,V\,c}},$
where $\varepsilon$  is the scalar lattice deformation, 
$\rho$ and $V$ the crystal mass-density
and volume, while $c$ is the sound velocity.

Let us now focus on the explicit form of the function $\Gamma$ in 
(\ref{explicit}, i.e.,
\begin{eqnarray}
& &\Gamma_{ij}^{\pm}=2\,\pi\sum_{\bf{q}} g_i({\bf{q}})\, \bar g_j({\bf{q}})\,
(n_{\bf{q}}+\theta(\pm))\,\delta(\omega_{\bf{q}}-\omega)\nonumber =\\
& & \frac{V}{(2\pi)^2} \int {d^3{\bf q}}\, 
 \,g_i({\bf{q}}) \bar g_j({\bf{q}})\, (n_q+\theta(\pm))\delta
(\hbar\,c\,q-\hbar\,\omega)
\label{Gamma}
\end{eqnarray}
Thanks the axial symmetry of the problem and the delta function of energy 
conservation, 
the three-dimensional integral  over ${\bf q}$ in (\ref{Gamma})
is better approached in polar coordinates: 
$d^3{\bf q}={ q}^2\,d\varphi d(\cos \vartheta) d q$. 
One obtains an expression proportional to   
 \begin{eqnarray}
 \int_{-1}^{1} dt\, e^{\frac{Q^2\, t^2} {4\,a}}
\frac{ \cos(\,Q t\,z_{ij})}{[(q/Q)^2-t^2]^2}
\frac{1-t^2}{t^2} \sin^2[\frac{\pi t}{q/Q}].
\label{matr-el}
\end{eqnarray}
with $q=q_0,\,a=a_0,\,Q=E/\hbar c$. Moreover,
 $z_{ij}=a\,(i-j)$ is the distance between $i$-th and $j$-th QD's.
The crucial point is to observe that, for $Q/a_0^{1/2}= 
\lambda_\perp/\lambda_\parallel,\,(\lambda_\parallel\sim
Q^{-1}$) large enough, 
this integral
is dominated by contributions around $t\equiv\cos \vartheta=1$; 
therefore  
\begin{equation}
\Gamma_{ij}^{(\pm)} = \Gamma_{11}^{(\pm)}\, \cos[Q\,z_{ij}].
\end{equation}
Recalling that $\lambda_\perp=a_0^{-1/2}$ is the typical 
length scale of carrier confinement 
in the $x$-$y$ plane, this behaviour is easily understood:
due to the energy-conservation constraints 
(${\bf q}_\perp^2+ { q}_z^2
=|{\bf q}|^2=Q^2$),   
for delocalized in-plane wavefunctions
(with respect to the length scale $\lambda_\parallel,$)
the significant fluctuation of ${\bf q}$ in the considered state is small; 
therefore 
$q_z\simeq Q.$
In other words, due to  the exponential suppression ---in the overlap 
integral---
of the contributions from phononic modes with non-vanishing in-plane
components {\em the system behaves 
as in the presence of a single effective phonon mode
along the $z$ axis resonant with the qubit excitations}. 
As clearly confirmed by our numerical analysis reported in Sect.~\ref{s:Sim},
this is an extremely important feature of the semiconductor model 
considered: 
in spite of its 3d nature
and of the presence of a continuum of decoherence-inducing phonon modes,
in this regime the carrier subsystem experiences
an {\em effectively $1$-d  coherent environment}, 
that in a good approximation can be described by the Circular Model (CM)
analysed in App.~A.\cite{notCM}

This model, parametrized by the dimensionless quantity $\tilde Q\equiv Q\,a$,
represents a non trivial example of a register-environment coupling that admits a rich
structure (as a function of  $\tilde Q$) of sub-decoherent encodings.
From this point of view, it realizes a generalization of the replica symmetric model
(pure collective decoherence) discussed in \onlinecite{ZR}, 
that is recovered for $\tilde Q=0$.
Here, we limit ourselves to summarize the main result:

Safe quantum encoding are  possible
for the models such that $e^{i\,\tilde Q}$ is  a $4$-th roots of the unity,
the  most efficient case being the  points
$\tilde Q=0,\,{\em mod}\,\pi$;
when all the register cells feel the same external coupling
the dynamics is maximally collective thanks the full permutational
symmetry.

The existence of infinitely many ``magic'' points
is clearly due to the unphysical nature of the CM
that allows for undamped interactions between objects separated
by arbitrary large distances.
In realistic systems
(as the ones investigated in this paper)
the cosine dependence of the $\Gamma$ matrix can be only approximated
and the periodicity with respect to the cell distance
eventually destroyed by some overimposed decay.
In a way, the present situation is very similar to having a string of 
(two-level) atoms
in a cavity coupled with a single resonant electromagnetic  mode 
\cite{GHE}. 

\section{Simulation of sub-decoherent dynamics in a QD array}\label{s:Sim}

In this section we will present our numerical analysis of subdecoherent 
quantum encoding for realistic QD structures. 

\subsection{Carrier-Phonon Scattering in a Single QD Structure}

As a starting point, let us discuss the role of carrier-phonon interaction 
in a single QD structure. 
Figure \ref{fig1} shows the total (emission plus absorption) carrier-phonon 
scattering rate at low temperature ($T = 10$\,K)
as a function of the energy spacing $E$ for three different values of the 
GaAs QW width ($d = 3, 4, and 5$\,nm).
Since the energy range considered is smaller than the optical-phonon energy
 ($36$\,meV in GaAs ), due to energy conservation scattering with LO phonons
is not allowed. Therefore, the only phonon mode
$\lambda$ which 
contributes to the rate of Fig.~\ref{fig1} is that of acoustic 
phonons. Again, due to energy conservation, the only phonon wavevectors 
involved must satisfy
$|{\bf q}| = {E/\hbar c_s}\equiv q$, 
$c_s$ being the GaAs sound velocity. It follows that by increasing the 
energy spacing  $E$ the wavevector $q$ is increased, 
which reduces the carrier-phonon 
coupling entering in the electron-phonon
interaction and then  the scattering rate. 
This well-established behaviour, known as phonon bottleneck,\cite{BN} 
is typical of a quasi-0D structure. 
As shown in Fig.~\ref{fig1}, for 
$E = 5$\,meV ---a standard value for many state-of-the-art QD structures---
the carrier-phonon scattering rate is 
already suppressed by almost three orders of magnitude compared to the 
corresponding bulk values.\cite{Shah,Kuhn} 
 
In addition to the bottleneck scenario discussed so far, for a given value 
of the energy spacing $E$ we see that for small values of $d$ we have an 
increase of the carrier-phonon rate. In spite of the reduction of the 3D 
volume available to the carrier states, the overall coupling is increased, 
basically due to the progressive relaxation of momentum conservation along 
the growth ($z$) direction.

\subsection{Short-Time analysis}

We will now show that by means of a proper information encoding, i.e., a 
proper choice of the initial multi-system quantum state, and a proper design
of our QD array,
we can strongly suppress phonon-induced decoherence processes, thus further 
improving the 
above single-dot scenario.
To this end, let us consider a four-QD array, 
which is the simplest noiseless qubit register
(see App.~A) 
From the short-time 
expansion discussed in Sect.~\ref{s:ME}, 
we have numerically evaluated the decoherence 
rate for such QD array choosing as energy splitting $E = 5$\,meV and QW 
width $d = 4$\,nm (see Fig.~\ref{fig1}). 
As  initial state we have chosen 
the singlet $|\psi_{{\cal D}_1,{\bf{0}}}\rangle$ [see Eq. ~\ref{States}] 
defined by the dimer
partition ${\cal D}_1=\{(1,2),\,(3,4)\}$.
We stress that, when the CM approximation (see App.~A) 
is not exactly fulfilled,
different singlets have different decoherence rates.
Indeed, the larger is the distance 
$z_{ij}$ between the pair elements in the dimer covering,
the greater is the deviation from the strictly periodic behaviour.
Thus from Eq. (~\ref{rate}) it follows, for instance,   
that the singlet corresponding to the dimer partition
${\cal D}_2=\{(1,3)\,(2,\,4)\}$ has a greater decoherence rate than 
$|\psi_{{\cal D}_1,{\bf{0}}}\rangle$;
The decoherence rate obtained from our numerical calculation 
is shown as solid line in Fig.~\ref{fig2}(a) as 
a function of the inter-dot distance $a$. 
The uncorrelated-dot decoherence rate is
also reported as dashed line for comparison.
As suggested by the analysis of the circular model 
presented in App.~A, 
in spite of the 3D nature of the sum over ${\bf q}$ 
entering the calculation of the function $\Gamma^{(\pm)}_{ii'}$ 
[see Eq.~(\ref{matr-el})], 
the decoherence rate exhibits a periodic behaviour
over a range comparable to the typical
QD length scale.
In the circular-model approximation (and for $T=0$) one obtains
$\tau_1^{-1}[|\psi_{{\cal D}_1}]\simeq 2\,\Gamma_{00}^{(-)}\,[1-\cos(Q\,a)],$
from which it follows that for $a_n= 2\,n\,\pi/Q,\,(n\in{\bf{n}}$
the considered state is stable. 
This effect  -- which 
would be natural for a 1D phonon 
system -- 
 stems from the exponential suppression, in the overlap integral,
of the contributions of phononic modes with non-vanishing in-plane 
component, previously discussed. 
This 1D  behavior 
is extremely important since it allows, by suitable choice of
the inter-dot distance $a$, to realize a symmetric regime 
in which all the dots experience the {\sl same} phonon field and therefore 
decohere collectively. 
Figure \ref{fig2}(a) shows  that for the particular QD structure considered,
case C should correspond to a decoherence-free evolution of
a singlet state, which is not the case for A and B (see symbols
in the figure).

In order to better understand how this sort of effective 1D behaviour 
depends on 
the material parameters considered, we have repeated the subdecoherence 
analysis of Fig.~\ref{fig2}(a) by artificially increasing the GaAs 
effective mass. More specifically, Figs.~\ref{fig2}(b) and \ref{fig2}(c) 
present the same decoherence analysis, respectively, for values of $5$ and 
$10 m^*$.
As we can see, by increasing the effective-mass value the 1D character in (a)
is progressively suppressed. This can be clearly understood as follows: the
increase of the effective mass leads to a stronger and stronger 
localization of the 2D harmonic-oscillator wavefunctions which, in turn, 
can easily interact with transverse ($xy$) phonon modes ${\bf q}$.

As far as the unitary component of the Liouvillian is concerned,
one can easily show that [for any $|\psi\rangle$ eigenstate of $S^z$]
$F(t)= |\langle\psi |\, e^{-i\,t\, H_{\cal R}^\prime}\,|\psi\rangle|^2 = 
1-(t/\tau_U)^2+ o(t^3),$  where 
\begin{equation}
\frac{2}{\tau_U^2}=\langle\psi|\,\delta H_{\cal R}^2\,|\psi\rangle-
\langle\psi|\,\delta H_{\cal R}\,|\psi\rangle^2. 
\end{equation}
Figure \ref{fig3} shows $\tau_U^{-1}[|\psi_{{\cal D}_1}]$ as a function 
of the inter-dot distance $a$. 

We find an oscillatory behaviour similar to that of Fig.~\ref{fig2}(a); 
it stems from the fact that (for the material parameters considered)
$\Delta^{\pm}_{ij}\simeq \Gamma_{00}^\pm \sin[ Q(i-j)a +\varphi],$
with $\varphi\ll\pi/2$. 
Thus, for values of $a$ corresponding to a subdecoherent dynamics [see 
point C in Fig.~\ref{fig2}(a)] 
the $\Delta$ contribution, also known as polaronic 
shift, is negligible as well.

\subsection{Time-dependent solution of the Master Equation}

In order to extend the above short-time analysis, we have performed 
a direct numerical
integration of the   Master equation (see Sect.~\ref{s:ME}), 
thus obtaining the reduced density matrix $\rho$ as a function of time.
Also the Lamb-shift terms discussed in
Sect.~\ref{s:Theory}
have been taken  into account.
Starting from the same GaAs QD structure considered so far, we have 
simulated the above noiseless encoding for a four-QD array. 
Figure \ref{fig4} shows the fidelity as a 
function of time as obtained from our numerical solution of the Master
equation. In particular, we have performed three different simulations 
---for the same initial state
$|\psi_{ {\cal D}_1, {\bf{0}}}\rangle$--- 
corresponding to the different values of $a$ depicted in Fig.~\ref{fig2}(a).
Consistently  with our short-time analysis, 
for case C we find a strong suppression of the 
decoherence rate which extends the sub-nanosecond time-scale of the  B 
case (corresponding to the uncorrelated-dot rate) to the microsecond time-scale.

An other quantity which properly describes the environment-induced 
corruption of information is 
 the linear entropy
$\delta[\rho]\equiv \mbox{tr} (\rho-\rho^2)$.  
Its production rate is also directly connected to $\tau_1$; 
indeed for an initial pure preparation we have
$\dot \delta(t) =2\,t/\tau_1 +o(t^2)$. 
for intial pure preparations.
The time evolution 
of the linear entropy, as obtained from our numerical solution of the ME, 
is reported in Fig.~\ref{fig5}.
We can clearly recognize an initial transient (of the order of $\tau_1$)
in which the register, getting entangled with the environment, 
decoheres;
this is followed by a  subsequent  slower relaxation dynamics.

The time-dependent analysis of Figs.~\ref{fig4} and \ref{fig5}
confirms that by means of the proposed encoding strategy one can realize a 
decoherence-free evolution over a time-scale comparable with typical 
recombination times in semiconductor materials \cite{Shah}.\\

\section{Summary and conclusions}\label{s:Conc}

We have investigated a possible semiconductor-based implementation 
of the subdecoherent quantum-encoding strategy, i.e. error avoiding, 
recently proposed in \onlinecite{ZR}.
The goal is the suppression
of phase-breaking processes in a quantum register realized
by the lowest energy  {\em charge} excitations of a semiconductor 
QD array.\cite{ZARO}
In this case, the primary noise source is given by 
electron-phonon scattering, which is considered  to be the most
efficient    decoherence channel in such a system.\cite{Shah,Kuhn}

The main result is that, in spite of the 3D nature of carrier-phonon 
interaction in our QD structure,
by  means of a proper quantum encoding as well as of a proper tailoring of 
the semiconductor structure,
one can in principle increase
the coherence time by several orders of magnitude with respect to the 
bulk value. This would allow to realize a coherent quantum-mechanical evolution 
on a time-scale longer compared to that of ultrafast optical spectroscopy.
From this point of view this result might constitute 
an important  step toward
a solid-state implementation of quantum computers. 
On the other hand, it certainly represents a first non-trivial example 
of a solid-state quantum system for which one can apply
quantum  error avoiding strategies.

As already discussed in Sect. \ref{s:Appl}
 carrier-phonon scattering is not 
the only source of 
decoherence in semiconductors. In conventional bulk materials 
also carrier-carrier interaction is found to play a crucial role. However, 
state-of-the-art QD structures ---often referred to as semiconductor 
macroatoms~\cite{QD-reviews}--- 
can be regarded as few-electron systems basically decoupled from the 
electronic degrees of freedom of the environment. 
For the semiconductor QD array considered,
the main source of Coulomb-induced ``noise'' may arise from the inter-dot 
coupling. However, since such Coulomb coupling vanishes for large values of 
the QD separation and since the proposed encoding scheme 
can be realized for values of $a$ much larger than the typical 
Coulomb-correlation length (see Fig.~\ref{fig2}), 
a proper design of 
our quantum register may  rule out 
 such additional decoherence channels.\cite{notPH}

The actual implementation of the suggested encoding relies, of course,
on precise quantum state synthesis and manipulations.
This further step, not addressed in this paper,
represents the most challenging open issue concerning the ultimate
usefulness of the proposed coding strategy.

\section*{Acknowledgments}

We are grateful to M. Rasetti  for stimulating and fruitful discussions. 
This work was supported in part by the EC Commission through the TMR Network 
``ULTRAFAST''. P.Z. thanks Elsag-Bailey for financial support.
\begin{figure}
\caption{\label{fig1}
Carrier-phonon scattering rate for a single QD structure as a function of
the energy splitting $E$ for different values of the QW width $d$ at low
temperature (see text).
}
\end{figure}
\begin{figure}
\caption{\label{fig2}
(a) Phonon-induced decoherence rate
for a four-QD array (solid line) as a
function of the inter-dot distance $a$ compared with the corresponding
uncorrelated-dot rate (dashed line);
(b) Same as in (a) but with an artificial effective mass of $5m^*$;
(c) Same as in (a) but with an artificial effective mass of $10m^*$
 (see text).
}
\end{figure}
\begin{figure}
\caption{\label{fig3}
$\tau_U^{-1}[|\psi_{{\cal D}_1}\rangle]$ as a function of $a$ (see text).
}
\end{figure}
\begin{figure}
\caption{\label{fig4}
Fidelity $F$ as a function of time as obtained from
a direct numerical solution of the Master equation for the relevant case of
a four-QD array (see text).
}
\end{figure}
\begin{figure}
\caption{\label{fig5}
Linear entropy as a function of time as obtained from
a direct numerical solution of the Master equation for the relevant case of
a four-QD array (see text).
}
\end{figure}


\appendix\label{app}
\section{Circular-interaction model}

This  appendix is devoted to the formal analysis of a 
 model
with periodic (environment-induced) interactions between register cells.
We set $\Gamma_{ij}^{(\pm)}=\Gamma^{(\pm)}\cos [Q\,(i-j)]$;
the resulting model will be referred to as the {\em circular model} (CM).
The dimensionless parameter $Q$ is taken to be given by the product
of a characteristic wave vector 
(corresponding to an effective one-phonon field) times
the inter-cell distance. 
The effective Hamiltonian (\ref{Ham}) takes  the form $\tilde H=  \sum_{\alpha=\pm}
H^{(\alpha)}(Q),$ with
\begin{equation}
 H^{(\alpha)}_Q= \frac{1}{2}{\Gamma^{(\alpha)}}
( S^{-\alpha}_Q\,S^{\alpha}_{-Q}+S^{-\alpha}_{-Q}\,S^{\alpha}_{Q})
\label{HamQ}
\end{equation}
where
$ S^{\alpha}_Q=\sum_{j=1}^N e^{i\,Q\,j} \sigma^\alpha_j\;(\alpha=\pm,\,z),$
are the present Lindblad operators.
 They fulfill the  following commutation relations 
\begin{eqnarray}
& &[ S^{\pm}_Q,\,  S^{\mp}_{ Q^\prime}]= \pm 2\, S^z_{Q+Q^\prime}\nonumber \\
& &[S^z,\, S^{\pm}_Q]= \pm\, S^{\pm}_Q,\quad
\label{LieGrade}
\end{eqnarray}
For $Q=0\,\mbox{mod}\,2\,\pi$ one recovers 
the  global $sl(2)$ algebra spanned by the $S^\alpha$'s,
 to which the $S^\alpha_Q$'s are connected by the following 
 unitary transformations
$U_Q\equiv\exp( i\,Q\,\sum_{j=1}^N j\,\sigma_j^z).$
Indeed, we have 
$S^\alpha_Q=U_{\alpha Q} \,S^\alpha U_{\alpha Q}^\dagger\,(\alpha=\pm)$
 (notice that $U_Q^\dagger =U_{-Q}$).
In terms of these unitary transformations and of the $Q=0$ Hamiltonian 
$H_0=\Gamma^{(-)}
\, S^+\,S^- + \Gamma^{(-)}\,S^-\,S^+$
the CM model (\ref{HamQ}) reads
\begin{equation}
 H_Q= 2^{-1}\sum_{\eta=\pm} U_{\eta Q}\,H_0\,U^\dagger_{\eta Q}
\label{HamQ1}
\end{equation}
From Eq.~(\ref{LieGrade}) it follows that, for {\em any generic} $Q$,
the two terms in the above equation do not commute:
the model is  non trivial, i.e., non integrable.

Next proposition  shows that
the analytic structure 
of the CM strongly depends on the input parameter $Q,$
for particular $Q$ values it is  quite simple and its subdecoherent 
coding efficiency is optimal. 

{\bf Proposition  1} {\em  One has the following integrable points  
\begin{itemize}
\item[i) ] $Q=0\, \mbox{\em mod}\, 2\,\pi,\Rightarrow H^{(\alpha)}(2\,\pi) =\Gamma^{(\alpha)} 
S^{-\alpha}\,S^{\alpha},$ replica symmetry.
\item[ii)]  $Q=\pi\, \mbox{\em mod}\, 2\,\pi,\Rightarrow H^{(\alpha)}(\pi) =\Gamma^{(\alpha)}
S^{-\alpha}(\pi)\,S^{\alpha}(\pi)= U_\pi\,H^{(\alpha)}(2\,\pi)\,U_\pi^\dagger.$
\item[iii)] if $Q=\pi/2, 3/2\,\pi \, \mbox{\em mod}\, 2\,\pi$ one has
$\Gamma^{(\alpha)}_{i,i+2\,n}=\Gamma^{(\alpha)} (-1)^n$ and 
$\Gamma^{(\alpha)}_{i,i+2\,n+1}=0.$
The odd- and even-site sublattices decouple, and for each sublattice
case ii) is recovered.
\end{itemize}}

Notice that for cases i) and ii) 
 $2\,Q=0\, \mbox{mod}\, 2\,\pi$; 
then the $\eta=+$ and $\eta=-$ terms in (\ref{HamQ1}) are identical;
the model is then unitarily equivalent to the $Q=0$ case.
The latter is clearly diagonalized in the $S^2,\,S^z$ eigenbasis and its   
spectrum is given by
$E=\sum_{\alpha=\pm} E^{(\alpha)}(J,\,M,\,r)$ where
\begin{equation}
 E^{(\alpha)}(J,\,M,\,r)=\Gamma^{(\alpha)}\,[J\,(J+1)-M\,(M+\alpha)],
\end{equation}
$J=J_{min},\ldots,N/2;\, M=-J,\ldots,J;\,r=1,\ldots,n(J,N),$
in which $J_{min}=0$ ( $J_{min}=1$) for $N$ even (odd), and $n(J,N)$ 
denotes the multiplicity of the $sl(2)$ representation 
labelled by $J$~\cite{ZR}
 \begin{equation}
n(J,N)=\frac{N!\,(2\,J+1)}{ (N/2+J+1)!\,(N/2-J)!}. 
\label{mult}
\end{equation}
If $N$ is even and $0 < \Gamma^{(+)}\le \Gamma^{(-)}$ 
the lowest eigenvalue
is $E_0=0$ with degeneracy $n(0,\,N),$ the ground-state manifold being the
{\sl singlet} sector of the global $sl(2).$
At zero temperature one has $\Gamma^{(+)}=0$;
therefore all the lowest-weight $sl(2)$-vectors $|J,\,-J\rangle$
are ground states of $\tilde H.$ 

Let us consider the $N$-th roots of the unit (with $N$ even)
\begin{equation}
{\cal Z}_N=\{ e^{i\,Q_j}\,/\, Q_j=\frac{2\,\pi\,j}{N},\, j=0,\ldots,N-1\}.
\label{Zn}
\end{equation}
This (multiplicative) group is of course isomorphic to the 
(additive) group ${{\bf Z}}/{N\,{\bf Z}}=\{0,\ldots,\,N-1
\}$; 
thus we shall use the same notation for both.
 Here
${\cal Z}_N$ is considered a subgroup of ${\cal S}_N.$
The latter as a natural action on ${\cal H}_{\cal R}$ given by the 
linear extension
$p\colon \otimes_{j=1}^N|\sigma_j\rangle
\mapsto\otimes_{j=1}^N|\sigma_{p(j)}\rangle,\,( p\in {\cal S}_N).$ 

The operators $S^{\alpha}_m\equiv S^\alpha(Q_m),$
satisfy to the commutation relations $[S^\alpha_m,\,S^\beta_n,]= K^{\alpha\beta}_\gamma\, S^\gamma_{n+m},$
$K^{\alpha\beta}_\gamma$ are the $sl(2)$ structure constants.
They  span the (${\cal Z}_N$-graded) Lie  algebra 
\begin{equation}
{\cal A}_N\equiv\mbox{span} \{ S^{\alpha}_m\,/\, \alpha=z,\,\pm,\;m\in{\cal Z}_N\}
\cong \oplus_{i}^N sl(2)_i.
\end{equation}

Let ${\cal A}_N^{Q}$ the Lindblad operators algebra for a generic $Q,$
the following proposition gives a characterization of it  when  $Q$  varies.

{\bf Proposition 2} 
\begin{itemize}
\item[i)] 
{\em For a generic $Q$ (i.e., $e^{i\,Q}\notin {\cal Z}_N$) one has 
${\cal A}_N^{Q}\cong {\cal A}_N,$
whereas for $e^{i\,Q}\in {\cal Z}_N$ 
one finds} 
$$
{\cal A}_N^{Q_n}=\mbox{span}\{S^z_{2\,p\,n},\,S^{\pm}_{n\,(2\,p+1)}\,/\,p\in{\cal Z}_N\}
$$
\item[ii)] ${\cal A}_N^{0}\cong {\cal A}_N^{\pi}
\cong sl(2)$ 
\item[iii)] ${\cal A}_N^{\pi/2}=sl(2)_e\oplus\, i\, sl(2)_o,$ where
\begin{eqnarray}
sl(2)_e &\equiv& \mbox{span}\{ \sum_{j=1}^{N/2}(-1)^j \sigma^\alpha_{2\,j}\}_{\alpha},
\nonumber \\
sl(2)_o &\equiv& \mbox{span}\{ \sum_{j=0}^{N/2-1}(-1)^j \sigma^\alpha_{2\,j+1}\}_{\alpha}
\end{eqnarray}
\item [iv)] $e^{i\,Q_j}\in {\cal Z}_N^*\equiv {\cal Z}_N-{\cal Z}_4\Rightarrow \mbox{dim}\,{\cal A}_N^{Q_j}= 3\,N/2$
\end{itemize}
{\em Proof}\\
One can check that ${\cal A}_N^{Q}=\mbox{span}\{S^z_{2\,p\,Q},\,S^{\pm}_{Q\,(2\,p+1)}\,/\,p\in{\cal Z}\}$
if $Q$ is rationally independent from $2\,\pi$ the numbers $e^{i\,2\,p\,Q},\,e^{i\,Q\,(2\,p+1)}$
densely fill the unit circle, from which ${\cal A}_N\subset {\cal A}_N^{Q}$ \cite {not3}. 
Points ii)-iii) follow from prop. 1, and iv) can be checked 
by a simple calculation
$\hfill\Box$. 

Notice that $e^{i\,\pi/p}\in{\cal Z}_N$ iff $N=0\,\mbox{mod}\, 2\,p\,(p=1,2) .$
Remembering that $|\psi\rangle\in\mbox{Ker}\,H_Q\Leftrightarrow |\psi\rangle$
is annihilated by {\em all} the generators of ${\cal A}_N^Q$ 
---and then that the smaller is the algebra the greater is the code--- 
Proposition 2 seems to indicate that the ``magic'' $Q$'s possibly
relevant for subdecoherent encoding are just the ones such that 
$e^{i\,Q}\in{\cal Z}_N.$

Let us now consider the $Q$ dependence of the symmetry structure of our model.

{\bf Lemma} 
{\em Let ${\cal G}_Q$ the (maximal) symmetry group of $H_Q$, one has:}\\
i) $
{\cal G}_0\cong {\cal G}_\pi= {\cal S}_N,$ ii)
${\cal G}_{\pi/2}\cong {\cal G}_{3\,\pi/2}=  
{\cal S}_{N/2}\times {\cal S}_{N/2},$ iii)
for $e^{i\,Q}\in S^1 -{\cal Z}_4$ one has
${\cal G}_Q= {\cal Z}_N.$

Pictorially one can say that in the CM the register
has a regular polygon topology that for the special points
$Q=0,\,\pi$ ($Q=\pi/2,\,3/2\pi$) collapses to a point (dimer)
gaining  in this way a larger permutational symmetry.
This dynamical clustering is  associated with a greater sub-decoherent 
coding efficiency.\cite{ZAme}
Next proposition summarizes in a formal manner 
the present situation.

{\bf Proposition 3} {\em Let  $N$ even then   
\begin{itemize}
\item[i)]  $e^{i\,Q}\in{\cal Z}_N\Leftrightarrow d_N(Q) > 0$ 
\item[ii)] $d_N(Q_0)=d_N(Q_{N/2})=n(0,\,N).$ 
\item[iii)]If $N=0\,{\em mod}\,4\Rightarrow 
d_N(Q_{N/4})=d_{N/2}(Q_{0})^2 $
\item[iv)] $e^{i\,Q_j}\in {\cal Z}_N^*\Rightarrow d_N(Q_j)=1.$
\item[v)] When $ e^{i\,Q_j}\in {\cal Z}_N^*$ the null space is spanned by the vectors
\begin{equation}
|\psi_j\rangle= \otimes_{i=1}^{N/2} (|01\rangle-(-1)^j\,|10\rangle)_{i,i+N/2}
\end{equation}
\item[vi)] { Let  $N$ odd then} $d_N(Q)=0\,\forall Q.$
\end{itemize}}
{\it Proof}\\
The cases $e^{i\,Q}=\pm 1$ are isomorphic and have been previously discussed.
Notice that, if  $H\ge 0$ one has
$\langle\psi |H|\psi\rangle=0 \Leftrightarrow  H\,|\psi\rangle=0.$
Moreover, $|\psi_j\rangle\equiv |\bar {\cal D},\,\gamma_j\rangle$
where $\bar {\cal D}$ is the unique dimer partition of the array with 
$|l-k|=N/2 $ and $\gamma_j(l,\,k)= j\,(\mbox{mod}\, 2)\,
\forall (l,k)\in\bar{\cal D}$.
From the second of Eqs.~(\ref{rate}) one finds that $$
\langle\psi_j | H_{Q_j}|\psi_j\rangle\sim1-\frac{2}{N}\sum_{l=1}^{N/2}(-1)^j\cos(\pi\,j)=0
$$
from which the sufficiency parts of i) and v) follow.
If $e^{i\,Q} \notin{\cal Z}_N$ from prop.  2 and 3, one has that
if $|\psi\rangle\in \mbox{Ker}\, H_Q$  
then it is in  the singlet sector of ${\cal A}_N^Q$ ( prop. 2)
 Since ${\bf C}^{2^n}$ is an irreducible (non-trivial)
representation
space of ${\cal A}_N$ such a sector is empty. 
Points ii)-iii) follow directly from prop. 2, 3.
Since the $S^\alpha_{\pm Q}$'s  transform 
according $1$-d ${\cal G}_Q$-irreps, from representation theory
it follows that 
$d_N(Q)$ (i.e., multiplicity of the $1$-d 
${\cal A}^Q_N$-irrep ) is equal to  
 the dimension of an irrep of the symmetry group 
${\cal G}_Q.$  But for $e^{i\,Q}\in {\cal Z}^*
_N$
one has ${\cal G}_Q\cong {\cal Z}_N,$ (abelian) therefore
its irreps are $1$-d, from which point iv) follows.\cite{not1d}
Finally, vi) simply stems from the fact that the necessary condition
$S^z\,|\psi\rangle=0$ cannot hold for odd  $N.$
$\hfill\Box$

To understand in a more constructive fashion, why the
$|\psi_j\rangle$ are the (only) subdecoherent states for $e^{i\,Q_j}\in{\cal Z}_N^*$
let us consider the following state $|\psi\rangle\in {\cal C}$
( $\cal C$ the (global) $sl(2)$-singlet sector) 
such that
i)  $U_{2\,Q}\,|\psi\rangle=|\psi\rangle.$
Then 
$U_Q\, |\psi\rangle= U_Q^\dagger\,|\psi\rangle=
U_{-Q}\,|\psi\rangle \equiv |\tilde \psi\rangle .$
This means $|\tilde \psi\rangle\in \bigcap_{\alpha=\pm} U_{\alpha\,Q}\,{\cal C}$
it follows that  $|\tilde \psi\rangle$ is annihilated by
$S_{\pm Q}^\alpha =U_{\pm Q}\,S^\alpha\,U_{\pm Q}^\dagger,\,(\alpha=z,\,\pm).$
and therefore by $H_Q$
It is now easy to check that the states $|\psi_j\rangle$ of prop. 4 are just
$U_{Q_j}\,|\bar {\cal D},{\bf 0}\rangle,$
the dimer partition $\bar {\cal D}$ being the only one allowing for
condition i) to be fulfilled.

\end{multicols}

\begin{references}
\bibitem{notDeco} The environment can  nevertheless affects the register
through the action of $\delta H_{\cal R}.$
\bibitem{notN2} The   CM Hamiltonian for $N=2$ can be written
in terms of global 
$S^\alpha$'s.
\bibitem{notWA}
This assumption is made for computational simplicity. 
Realistic (i.e., finite) barriers, would  give rise to inter-dot hopping.
Nevertheless, due to the exponential decay
of inter-dot overlap integrals, 
the effect is negligible also for rather small inter-dot distances.
\bibitem{note-degeneracy}
With ``first excited state'' here we mean any linear combination of the two
energetically degenerate states.
\bibitem{note-disp}
Notice that these matrices are related by a kind of Kramers-Kronig dispersion  relation.
This stems from the fact that they are basically real and imaginary part
of a single electron propagator obtained by tracing out the phonon field \cite{ZARO}.
\bibitem{notCM} It is worthwile to stress that this result appears to be largely
independent on the detailed shape of the qubit wavefunctions 
[see Eq.~\ref{matr-el}].
 Indeed, the only requirement that appears to be crucial
is the x-y delocalization, with respect to the effective phonon wavelength.
\bibitem{notPH} The reader may wonder whether the increase of $a$ 
makes the various qubits interacting with independent (i.e., uncorrelated) 
phonon baths.
We observe that a  lower bound for the acoustic-phonon life-time 
in GaAs is tenths of picoseconds,
which corresponds to a coherence length
much larger than the typical inter-dot distance $a$
relevant  for noiseless encoding. This  makes the phonon dynamics fully
coherent on the space scale of our QD register.
\end{references}
\end{document}